# Effects of green revolution led agricultural expansion on net ecosystem service values in India

Srikanta Sannigrahi[a*], Francesco Pilla[a], Qi Zhang[b], Suman Chakraborti[c], Ying Wang[d], Bidroha Basu[a], Arunima Sarkar Basu[a], P.K. Joshi[e], Saskia Keesstra[f, g], P.S. Roy [h], Paul. C. Sutton[i], Sandeep Bhatt[j], Shahid Rahmat[k], Shouvik Jha[l], Laishram Kanta Singh[m]

[a] School of Architecture, Planning and Environmental Policy, University College Dublin Richview, Clonskeagh, Dublin, D14 E099, Ireland.
[b] Frederick S. Pardee Center for the Study of the Longer-Range Future, Frederick S. Pardee School of Global Studies, Boston University, Boston, MA 02215, USA
[c] Center for the Study of Regional Development (CSRD), Jawaharlal Nehru University, New Delhi 110067, India
[d] School of Public Administration, China University of Geosciences, Wuhan 430074, China
[e] School of Environmental Sciences (SES), Jawaharlal Nehru University, New Delhi 110067, India
[f] Soil, Water and Land-use Team, Wageningen University and Research, Droevendaalsesteeg3, 6708PB Wageningen
[g] Civil, Surveying and Environmental Engineering, The University of Newcastle, Callaghan 2308, Australia.
[h] Innovation Systems for the Drylands (ISD), ICRISAT, Pathancheru, Hyderabad - 502 324, India
[i] Department of Geography and the Environment, University of Denver, 2050 East Iliff Avenue, Denver, CO  80208-0710
[j] Department of Earth Sciences, Indian Institute of Technology Roorkee - 247667, India
[k] Department of Architecture and Regional Planning, Indian Institute of Technology Kharagpur, India.
[l] Indian Centre for Climate and Societal Impacts Research (ICCSIR), Kachchh, Gujarat 370465, India
[m] Agriculture and Food Engineering Department, Indian Institute of Technology Kharagpur, Kharagpur 721 302, India

*Corresponding author: **Srikanta Sannigrahi**

E-mail: (Srikanta Sannigrahi*) : srikanta.sannigrahi@ucd.ie

**Abstract**

Ecosystem Services (ESs) are bundle of natural processes and functions that are essential for human well-being, subsistence, and livelihood. The expansion of cultivation and crop land, which is the backbone of Indian economy, is one of the main drivers of rapid Land Use land Cover (LULC) changes in India. To assess the impact of the 'Green Revolution (GR)' led agrarian expansion on the total ecosystem service values (ESVs), we first estimated the ESVs (Billion US$) from 1985 to 2005 for eight eco-regions in India using several value transfer approaches. Five explanatory factors, i.e., Total Crop Area (TCA), Crop Production (CP), Crop Yield (CY), Net Irrigated Area (NIA), and Cropping Intensity (CI) representing the cropping scenarios in country were used in constructing local Geographical Weighted Regression (GWR) model to explore the cumulative and individual effects on ESVs. A Multi-Layer Perceptron (MLP) based Artificial Neural Network (ANN) algorithm was employed to estimate the normalized importance of these explanatory factors. During the observation periods, cropland, forestland and water bodies have contributed the most and form a significant proportion (80% – 90%) of ESVs, followed by grassland, mangrove, wetland, and urban built-up. In all three years (1985, 1995, and 2005), among the nine ESs, the highest ESV accounts for water regulation, followed by soil formation and soil-water retention, biodiversity maintenance, waste treatment, climate regulation, and gas regulation. Among the five explanatory factors, TCA, NIA, CP showed strong positive association with ESVs, while the CI exhibited negative association. The study reveals strong association between GR led agricultural expansion and ESVs in India. This study recommends formulation of vigorous ecosystem management strategies and policies to preserve and maintain ecological integrity and flow of uninterrupted ESs for the improvement of natural ecosystems and human well-being.

**Keywords:** *Ecosystem service value; Land use change; Ecology; Value transfer; Green revolution; India*

# 1. Introduction

Ecosystem Services (ESs) refer to benefits (provisioning, such as food and water; regulating such as regulation of floods, drought, land degradation, and disease; supporting such as soil formation and nutrient cycling; and cultural such as recreational, spiritual, religious and other non-material) that humans freely gain from natural environment and ecosystems, and these add to the human well-being (Fisher et al., 2009; Costanza et al. 1997; Brat and Groot, 2012). Whereas, the term ecosystem service function (ESF) denotes bundles of ecological and ecosystem processes functioning at any given ecosystem irrespective of whether or not such processes contribute to human well-being and/or other living organisms (Odum, 1956, Brat and Groot, 2012). Ecosystem service values (ESVs) are a monetary value assigned to an ecosystem and its services to assess the impact of anthropogenic activities on different ecosystems (e.g., MA, 2005; Adekola et al., 2015). ESV is a comprehensive assessment and has proven to be an alternative appraisal between environment and human development for sustainable natural resource management (Braat and de Groot, 2012; Potschin and Haines-Young, 2013; Pandeya et al., 2016; Adekola et al., 2015). The growing importance of ESs provide and help in adjusting the cost benefit analysis by evaluating both the negative and positive effects of any human developmnet activities including land use and land cover (LULC) change on the natural environment and ecosystems.

In India, the agricultural ecosystem is providing various valuable ecosystem goods and services that are essential for human well-being and subsistence (Power, 2010; Swinton et al., 2007). The introduction of the Green Revolution (GR) helped farmers to become financially stable by increasing their farm production (David and Otsuka, 1994). The GA approach involved the use of modern technologies, including high yielding variety seeds, irrigation facilities, farm machinery, chemical fertilizers, and plant protection measures. The GR raised many people out of deprivation, increased economic growth and prevented that forest wetlands and other fragile lands to be converted into agricultural land (Spielman and Pandya-Lorch, 2010). Due to the GR in India, during 1967-68 and 1977-78, the entire cropping system of the country has changed. The impact of the GR tripled the agricultural production, while the cultivated land area increased only by 11% (Singh, 2000). In India, the GR has positively impacted the country's economy and changed the way of life (David and Otsuka, 1994).

Agriculture and associated activities are the backbone of the Indian economy. It contributes substantially to the national income which in turn determines the overall growth of the nation. The GR cultivation system transformed India from a food-deficient country to one of the leading agricultural countries in the world. The impact of the GR on multiple ESs over all of India has not been adequately documented and analysed before. The significance of this study lies in the quantification of the impact of the successful agricultural expansion has had on the ESs in India. This study highlighted the success of Indian agrarian ESs at the expense of forests and other natural areas and the land reclamation through irrigation programmes. Few study has focused on creating agricultural based sustainability indicators (Rao et al., 2018), reconstructing of long term landuse (Ramachandran et al., 2018). But, to best of our knowledge, no such study is attempted earlier to analyze the ES valuation in an agriculture country like India, where, the GR has significantly changed the country's production landscape. Therefore, this study made an effort to unequivocally quantify the loss of ESVs (Billion US$) from 1985 to 2005 using multiple value transfer approaches (Costanza, 97a,b (Costanza et al., 1997), Costanza, 2011 (Costanza et al., 2014), de Groot, 2012 (de Groot et al., 2012), and Xie, 2008 (Xie et al., 2008)) in India. Therefore, objectives of this study are (1) to estimate the ESVs of different ecoregions in India, (2) to analyze the impact of agricultural expansion on ESVs in India.

## 2. Materials and methods
### 2.1 Calculation of LULC dynamics

Time series LULC data was used to estimate per unit ESV of each ecosystem types for three reference years. The LULC data was derived from the study of Roy et al., (2015) that has produced a detailed LULC information for India with the finest spatial resolution available (based on 30m, converted to 100m) for national database and temporal interval (available for 10 years interval,1985- 1995-2005). An overall 94.4% classification accuracy was achieved for all LULC categories for 2005 (evaluated using ~12606 sample points). For more information on the classification scheme and satellite data used for the study, one can refer to Roy et al., (2015). In our study, we have reclassified the original 17 LULC classes into eight major biomes, i.e., forest land, cropland, urban built-up, grassland, fallow land, water bodies, mangrove and wetland, in order to adjust with Costanza et al., (2014) defined equivalent biomes (**Table. 1**). Since we detect some miscalculation for mangrove category in Sundarban region in the original dataset, we have further adjusted this class with our LULC classification for 2005. Using the raster to features conversion approach, the spatiotemporal dynamics and conversion of LULC categories for three decadal periods, i.e., 1985–1995, 1995–2005, and 1985–2005, were quantified. The spatiotemporal LULC change dynamics were assessed as follows:

$$\Delta LULC_k = \frac{LULC_{end} - LULC_{start}}{LULC_{start}} \times 100 \qquad (1)$$

Where $\Delta LULC_k$ are the changes in the area of a LULC type $k$, $LULC_{end}$ and $LULC_{start}$ are the area of each ecosystem types at the end and start of the time periods. A transpose matrix was developed to quantify the spatiotemporal changes of different LULC categories. The LULC categories for the start and end years were assigned a specific code to calculate the area transferred among classes between the two reference years.

*2.2 Calculation of Ecosystem Service Value (ESV)*

The simple benefit transfer approach proposed by Costanza et al., (1997) was employed in this study to estimate ESVs of each LULC categories as follows:

$$ESV_k = \sum_f A_k \times VC_{kf} \qquad (2)$$

$$ESV_f = \sum_k A_k \times VC_{kf} \qquad (3)$$

$$ESV = \sum_f \sum_k A_k \times VC_{kf} \qquad (4)$$

Where $ESV_k$ is ESV of each LULC category $k$, $ESV_f$ is ESV of each LULC category $f$, and $ESV$ indicates the total estimated ESV, $A_k$ refer to the area (ha) of each LULC type $k$, $VC_{kf}$ is the equivalent value coefficient (US \$ ha$^{-1}$ year$^{-1}$) of each LULC type $k$ and type $f$, respectively (Richmond et al., 2007; Kindu et al., 2016; Sannigrahi et al., 2018). After that, the changes of ESVs were calculated as follows:

$$\Delta ESV = \frac{ESV_{end} - ESV_{start}}{ESV_{start}} \times \frac{1}{t} \times 100 \qquad (5)$$

Where $\Delta ESV$ refers to the change of ESVs of a particular LULC type $i$, $ESV_{end}$ and $ESV_{start}$ exhibits ESV of the past and current years, respectively and $t$ represents the time period.

Additionally, the equivalent value (EV) factor approach proposed by Xie et al., 2008, was used to estimate the ESV of the key ESs of India (**Table. 2, 3**). A preliminary study was conducted to carefully select the key ESs which are the best suitable for Indian ecosystems. Hence, to optimize the uncertainties and biases involved in the valuation process, multiple valuation methods, i.e., Costanza et al., (1997); Xie et al., (2008), De Groot et al., (2012) were adopted in this study. Later on, grouping and categorization of the selected ESs were done to obtain the biome/LULC specific ESV of India. Finally, total nine ESs, i.e., food production and production of raw materials are included in the provisioning services; gas regulation, climate regulation, water regulation, and waste treatment are included in the regulating services; soil formation, conservation and retention, and biodiversity maintenance are included in the supporting services; and aesthetic, cultural and recreation is included in the cultural ES. The food production EV for cropland ecoregions were first estimated using the Costanza et al., 1997 and Xie et al., 2008 valuation approaches. Consequently, the EV of other ESs were retrieved from the cropland equivalent factor. In the cropland valuation process, Liu et al., (2010) and Xie et al., (2008) have proposed that the projected food production service could be 1/7$^{th}$ of the real food production. This approximation was used to estimate EV for multiple key ESs in India. Information on crop production, crop price, crop yield, net irrigated area, and cropping intensity of the major crops of India were extracted from Directorate of Economics and Statistics, Department of Agriculture, Cooperation, and Farmers Welfare, Govt. of India (http://eands.dacnet.nic.in/latest_20011.htm), Open Government Data Platform (OGD), Govt. of India (https://data.gov.in), Ministry of Statistics and Programme Implementation (http://www.mospi.gov.in). The average food production and crop price in India during 1985-2005 were valued as 1495.21 kg ha$^{-1}$ and 0.27 US\$ kg$^{-1}$ (1 US\$ = 44.3 INR in 2005), and subsequently, the ESV of cropland food production service was estimated (1×1495.21×0.266/7).

*2.3 Elasticity of ESV to Land Use Land Cover (LULC) change*

The coefficient of elasticity ($C_{ES}$) is a standard concept in economics that means to measure the sensitivity of responding variable in order to change one control variable (Song et al., 2017). It depends on the self-condition of systems, having considered a constant elasticity and composition of an ecosystem (Yan et al., 2016). However, the specified weights of resistance and resilience depend on the probability of external forces (land use modification and associated alternation) surpassing or not the self-adjustment capability of any given ecosystem (Liu et al., 2017). In our study, we have assessed spatiotemporal elasticity of ESV of different ecoregions by LULC changes that would be helpful to identify the most sensitive and disturbed ecoregions at any given ecosystem. The $C_{ES}$ is estimated as follows:

$$C_{ES} = \left( \frac{(ESV_j - ESV_i)}{ESV_i} \times 100 \right) / LCI \qquad (6)$$

$$LCI = \frac{\sum_{i=1}^{n}(LULC_{jk} - LULC_{ik})}{\sum_{i=1}^{n} A_k} \times \frac{1}{t} \times 100 \qquad (7)$$

Where $C_{ES}$ is the coefficient of elasticity, $ESV_j$ and $ESV_i$ is the ESV's of ending and starting time, $LCI$ is the LULC change intensity, $LULC_{jk}$ and $LULC_{ik}$ is the area of land use type k at the ending and starting time, $t$ is the research period, $A_k$ is the area of land use type $k$.

## *2.4 Estimating the relationship between ESV and cropping pattern*
### *2.4.1 Fitting Geographical Weighted Regression (GWR)*

The geographical weighted regression (GWR) method, an extension of conventional ordinary least square (OLS) method was used in this study for its capability to capture the spatial variation that helps to assess the spatial association, spatial non-stationarity, and coefficient of determination (local $R^2$) between explanatory and response variables (Fotheringham et al., 2002). We fitted the GWR model to show how spatial variation of cropping pattern determines the ESV pattern. Therefore, the spatial weight was described based on its proximity to the location of observation (Su et al., 2014). However, the weight estimate of GWR is always sensitive to the selection of the kernel size and bandwidth parameterization. In addition, the observation with high proximity to the location of neighbouring features exhibits more significant influence than that of the distant elements which exerted less influence on parameter estimation (Fotheringham et al., 2002; Su et al., 2014). Additionally, the improper (coarser) parameterization of bandwidth and kernel selection would generate a global relationship and spatial stationarity, while a local estimate of spatial association and spatial non-stationarity is produced when bandwidth was set too small (Zou et al., 2016; Su et al., 2014). In this study, an adaptive kernel type was chosen for model parameterisation. The basic GWR equation is:

$$y(u_i, v_i) = \beta_0(u,v) + \beta_1(u,v)x_1 + \varepsilon(u,v) \qquad (8)$$

Where $y$ is dependent variable (ESV); $\beta$ is the intercept; $\beta_1$ is the coefficient; $v_i, u_i$ is the coordinates of sample $i$; $x$ is the independent variables (TCA, CP, CY, NIA, and CI); $e$ is the error.

The require weight matrix can be retrieved as follows:

$$w_{ij} = \exp\left(\frac{-D^2_{ij}}{B^2}\right) \qquad (9)$$

Where $w_{ij}$ is the weight of sample $j$ for sample $i$; $B$ is the kernel bandwidth; $D_{ij}$ is the distance between the sample $i$ and $j$.

### *2.4.2 Artificial Neural Network (ANN) for estimating relative effects of the cropping factors on ESV*

Artificial Neural Network (ANN) was used as a machine learning algorithm that enable a system to predict human learning processes through establishing and strengthening of the internal self-adjustment linkage system (Were et al., 2015; Wen et al., 2014; Qiang and Lam, 2015)). The ANN algorithms can efficiently predict, classify, make a decision and solve new problems through the trained parameters when the information is less. An ANN architecture is comprised of an input layer, a set of hidden nodes, and an output layer which is connected by a number of neurons (Chakraborti et al., 2018). In this study, we have adopted MLP neural networks with a backpropagation algorithm to predict and simulate the ESV pattern. In this network 30 hidden layers were chosen to generate optimum weights for predicting ESV, wherein 70%, 15%, and 15% samples were approximated for training, testing, and validating the model estimates. Additionally, we have performed simple and multiple linear regression (MLR) analysis to examine the single and joint effects of the explanatory factors on ESVs.

## 3. Results and discussion
### *3.1 Land use land cover changes in India during 1985 - 2005*

The spatial distribution of different LULCs for 2005 is represented in **Fig. 1**. Cropland areas are mainly distributed along the Indo-Gangetic Plain, Godavari, Krishna, and Cauvery basins, and part of Narmada, Tapi, and Mahanadi basins (**Fig. 1**). The highest proportion of forest cover in India is found in Central India, Eastern Himalayan region (North East India), and part of Gujarat, and in a scattered in Uttarakhand, and Himachal Pradesh.

The conversions of LULC were reported for three research periods, i.e., 1985–1995, 1995–2005, and 1985–2005 (**Fig. S1**). Between 1985 and 1995, the increasing trend of cropland areas at the expense of fallow land and forest land were documented predominantly in the western parts (Rajasthan, Gujarat) of India. The incentives for comprehensive watershed management and sustainable irrigation management practices for this arid region had stimulated agricultural productivity, and hence the cultivated agricultural areas increased dramatically (Davidar et al., 2010). The destruction of evergreen pine and deciduous broadleaf forest areas, especially, in the part of Odisha, North East India, and over the Western Himalayan states have also been documented in this period. These damages can be attributed to natural causes (landslide, wildfire, climatic anomalies) and human appropriations (deforestation, shifting cultivation, grazing, and human-made fire). In addition, a study by Meiyappan et al., (2017) on country scale land use change dynamics in India has observed that the proportion of irrigated cropland is negatively associated with the gross forest loss, which denotes the improvements of irrigation practices that can boost the cropping intensities of small, medium, and large farm sizes might would have minimised the pressure of forest conversion. This study (Meiyappan et al., 2017) has also reported strong linkages between the forest degradation and type of livelihood of the village communities in the area. The following activities are listed as being responsible for forest degradation in India (1) extraction of fuel woods, forest residuals and biomass products including wooden furnitures, timber products; (2) livestock cultivation including cattle, dairy, leather products; (3) villagers involving mining and quarrying activities; and (4) industrial set-up and development nearby forest landscapes. A study by Davidar et al., (2010) found that the proportion of agricultural households are negatively correlated with forest degradation. This could be due to the communities engaged with agricultural practices depend less on forest resources than communities that entirely depend on forest resources. During 1995–2005, substantial areas of cropland were reclaimed from fallow land, especially in the western parts of the country (comprising the dry areas of Gujarat, Rajasthan), and from forested and grassland regions in the southern (Tamil Nadu) part of India (FSI, 2003). Additionally, a exponential growth of urban built-up areas has been documented in this period. Research results have shown that during this period, significant areas of grassland were converted to forest cover in the Northern (Himachal Pradesh, Uttarakhand, Jammu and Kashmir), North-Eastern (Arunachal Pradesh, Assam, Meghalaya, Mizoram), Central (part of Madhya Pradesh), and South-Eastern (in a scattered way in Odisha) parts of India (**Fig. S1**). However, over the entire research periods (1985–2005), a net expansion of cropland and urban areas were documented at the expense of forest land, grassland, and fallow land respectively (**Fig. S1, Table 4**).

*3.2 Impact of land use changes on spatially explicit ESVs during 1985-2005*

Using the five unit values (Costanza, 1997a, Costanza, 1997b, Costanza, 2011, de Groot, 2012, and Xie, 2008), the mean ESV (Billion US$ year-1) of India was estimated for 1985, 1995 and 2005 (Fig. 2a, b). Forest and cropland ecosystems are providing the maximum (200–400 Billion US$ Year) ESs for all three reference years, with the maximum share (30–50%). Grassland, wetland, and waterbodies shared 5–15% of total ESVs for the given reference years. Except for the first reference periods (1985–1995), cropland ESV has increased throughout the research period (Fig. 2c). The maximum increase was observed during 1995–2005, followed by the 1985–2005 period. Whereas, the forest ESV has decreased substantially during the study period (Table. S1, S2).

The coefficient of elasticity of ESV to LULC changes are documented for three different time periods, i.e., 1985–1995, 1995–2005, and 1985–2005, respectively (**Fig. 3**). During 1985–1995, negative elasticities were documented for cropland, forestland, and mangrove eco-regions indicating the negative impact of LULC changes on ESV. During 1985–1995, the highest negative elasticity was observed for the forest eco-region indicating a negative impact of forest degradation and deforestation on country-level natural capital formation (**Fig. 3**). The negative elasticity resulted from any unwanted changes is reflecting the demeaning status of a particular ecosystem and seeks special attention and consideration for the improvement of natural resource management and preservation (Song, 2018; Sannigrahi et al., 2018). Considering the last half (1995–2005) and the whole research period (1985–2005), cropland eco-regions exhibit moderate to high elasticity to LULC changes (**Fig. 3**). The outcomes reveal the cumulative impact of agricultural expansion on total ESVs in India. However, the fast-tracked expansion of crop area and crop production during the research period and the resulted positive elasticity was found significantly lower than that of the negative elasticity of forest land. This indicates a higher capacity of natural forest ecosystems to produce green capital than any anthropogenic inputs (Costanza et al., 1997; 2014). Water bodies exhibited the second largest negative elasticity of ESV to the LULC changes, which is higher than cropland elasticity.

*3.3 Impact of the 'Green Revolution' on changing ESV patterns in India*

During the research period (1985–2005), the total estimated ESVs have increased mainly due to substantial expansion of cropland and wetland areas in India. Considering the total estimated ESV's in India, the cropland shares the major amount of mean ESV (30–40%) during the observation periods. This shows the enormous impact of agricultural productivity and subsequent food production on the green economy of India. India has a predominantly agriculture-based economy, which contributes almost 20–50% to the total national GDP. However, the contribution of agriculture-based economies on the formation of national GDP is gradually decreasing, as it accounted for 39% in 1983 and only 24% in 2000–2001. However, its contribution to total employment generation during the same period reduced only slightly (63–57%) (Mall et al., 2006).While the global ESVs show a decremental trend (Costanza et al., 2014; Sannigrahi et al., 2018a), the total estimated ESVs of India show an incremental tendency. The fast-tracked expansion of cropland areas in India, particularly during the period of 1995–2005 had happened due to (1) climatic favourability, normal to excess monsoon rainfall received during 1995 and 2005, as till now national average of 40% of the total cropped area in India is under the coverage of major and minor irrigation programmes, and almost 60% of the cultivated land is still rainfed and depends on seasonal monsoon rainfall (Mall et al., 2006; Guiteras, 2008; Roy et al., 2015). The two major cropping seasons of India, i.e., Kharif (June–September, entirely depends on summer monsoonal rainfall), and Rabi (October–November) provide the major productions of food grains and oilseeds of the country. The increasing trend of net primary production (NPP) of the country has also been documented during this period, aligned with the rainfall anomalies and resulting cropping pattern of India (Nayak et al., 2013); (2) Several major and minor irrigation projects like Indira Gandhi Canal System, Narmada Project and Accelerated Irrigation Benefit Programme (AIBP) were initiated in this period for boosting the crop production and resulting in an increase of net irrigated area. Additionally, this initiative has significantly increased the total cropped area at the expense of decreasing fallow land and forest land, especially in the western parts of India (Roy et al., 2015, **Fig. 4**). These programmes collectively increased the national irrigation potential of 5.44 million hectares under various major/medium irrigation projects and also generated 0.45 million hectares potential irrigation land under the multiple minor/small irrigation schemes up to 2009 (http://www.archive.india.gov.in/sectors/water_resources/index.php?id=8). Furthermore, the Ministry of Land Resource and the Ministry of Rural Development jointly adopted several area specific watershed management programmes: the 'Drought Prone Areas Programmes (DPAP)', the 'Desert Development Programme (DDP)', and the 'Integrated Wastelands Development Programme (IWDP)' to eradicate land degradation that successfully epitomizes the ecosystem as well as agricultural productivity (http://www.archive.india.gov.in/sectors/agriculture/index.php?id=7); (3) The area under plantation and aquaculture has increased substantially during the research periods (these LULC categories were merged into cropland types in this study, see **Table. 1**), especially in Southern India (Kerala, Tamil Nadu) and Western Himalayan region is also responsible for increasing observed cropland area in India (Roy et al., 2015). A significant forest ESV (9–19 Billion US$ year$^{-1}$) was lost during this period. This can be attributed to the substantial decrease of forest cover during the research period, specifically in the central and northeastern part of India (Roy et al., 2015). Different anthropogenic activities (biomass collections, including fuelwood, fodder, and green leaves harvesting by local communities), mining (including coal, iron, and aluminium ores), extensive shifting cultivation (especially in North-East India), population pressure and associated demand for agricultural land, construction of major dams and reservoirs; extraction of raw materials (cutting, burning, grazing, and re-cutting), and natural degradation (erosion, aggradation, landslides, wildfires, drought, climate change etc.) are the major reasons for depleting forest resources in India (Davidar et al., 2010; Munsi et al., 2010; Rao et al., 2015; Giri et al., 2011; Reddy et al., 2013; Roy et al., 2015; Semwal et al., 2004).

*3.4 Impact of cropping factors on ESVs in India*

**Fig. 5** shows the local estimates of GWR which demonstrate the total explained variance and predictive power of the explanatory variables (TCA, CP, CY, NIA, and CI) to estimate and predict ESV. Among the five explanatory variables, the TCA, CP, and NIA are highly associated with ESV compared to CY and CI for 1985, 1995, and 2005 (**Fig. 5**). In 1985 and 1995, the estimated ESV for Gujarat, Rajasthan, Haryana, Uttarakhand, and Uttar Pradesh were entirely dependent on TCA factors reflected by very high local $R^2$ approximation. High to moderate local $R^2$ was observed in Punjab, Himachal Pradesh, Maharashtra, Karnataka, Kerala, Tamilnadu, Chhattisgarh, Bihar and rest of the states of India. In 2005, a very high local $R^2$ was documented for the Central part of the country, due to the phenomenal increase of total crop area and resulting ESV (**Fig. 5**). Additionally, in 1985 and 1995, this study has proved that the CI factor does not have a significant impact on ESV, however, in 2005, the entire Northeastern states have produced a very low correlation between CI factor and ESV (**Fig. 5**). In addition, the CP factor has a notable impact on ESV. The entire Indo-Gangetic Plain regions and the Northeastern states characterized by high to a very high local $R^2$ approximation for 1985, 1995, and 2005. Whereas, the Central, Western, and Northern states of India exhibit low to moderate association between CP and ESV. Additionally, the CY factor shows a negligible to no coefficient of association with ESV (**Fig. 5**). Considering the total effects, the Central (Madhya Pradesh, Chhattisgarh), Eastern (West Bengal, Odisha, Tripura,

Mizoram, Manipur), and Southern (Andhra Pradesh, Tamil Nadu, Karnataka, Kerala, Telangana) states of India are earmarked by high to very high local $R^2$ during the research periods. While inspecting the normalized importance and weights of each input derived from the ANN approximation, the TCA factor was found to be the most important to predict ESV, followed by CP, CY, NIA, and CI, respectively (**Fig. 6**).

The linear effects of each explanatory factor on different ESs were examined and presented in **Fig. 7**. For gas regulation service, the highest coefficient of determination value was observed for TCA, followed by CY, NIA, CP, and CI, respectively. The TCA factor has attributed the highest coefficient of determination value for climate regulation service, followed by CP, CY, NIA, and CI. For water regulation service, the coefficient of determination values ranging from $R^2=0.59$ (TCA) to $R^2=0.002$ (CI) during the observation period. Concerning the soil formation and retention service, the TCA factor has explained the maximum model variances with high $R^2=0.66$ approximation, followed by NIA, CY, CP, and CI. For waste treatment service, the highest $R^2$ value was observed for TCA, followed by NIA, CP, CY, and CI. While accounting the model performances between the biodiversity maintenance service and explanatory factors, the TCA factor was able to explain the maximum model variances, followed by CY, NIA, CP, and CI. All the explanatory variables were performed most accurately with the least unexplained bias and estimates for food production service. The highest coefficient of determination was estimated for TCA, followed by NIA, CP, CY, and CI respectively. For raw material production service and recreation, culture, and aesthetic service, the percentage of model variances is ranging from 44% (TCA) to 0.2 % (CI) (**Fig. 7**).

Step-wise Multiple Linear Regression (MLR), along with ANOVA (F), and Student's t-test (t) was performed to examine the individual and cumulative effect of the explanatory variables, i.e. TCA, CP, CY, NIA, and CI, on ESV (**Table. 5**). Total 15 pairs of model were constructed to identify the best pair of model for predicting ESV. Among all models, model 1 explained the maximum model variances (85%) and found to be the best predictor of ESV with highest $R^2=0.85$, followed by model 2 ($R^2=0.45$), model 6 ($R^2=0.43$), model 13 ($R^2=0.43$), model 3 ($R^2=0.24$), model 10 ($R^2=0.19$), model 11 ($R^2=0.14$), model 14 ($R^2=0.12$). Among the explanatory factors, TCA (model 1) is exhibiting the most significant influence on ESV. Model 9 (P=0.24), model 12 (P=0.14), and model 15 (P=0.4) was found statistically insignificant in explaining model variances. This indicates the explanatory factors used for the model construction doesn't have any cumulative effects on ESV, except TCA. Model 10 and model 15 exhibited a negative correlation with ESV (**Table. 5**).

The pairwise correlation matrix was performed between the driving factors, and ESVs are shown in **Fig. 8**. All the pairs exhibited statistically significant correlation except CI factor. The regulating services are highly associated with other services and produced statistically significant estimates at P=> 0.001. A negative association was observed between CY and other factors, except CP. It can be seen in **Fig. 9** that all the driving factors except CI have produced significant association with the ESs. This exhibits that almost all the explanatory factors which reflect the GR led cropping scenarios in India have strong positive effects on the formation of natural capital and ES. After evaluating the individual effects of the driving factors on total ESVs, the TCA factor has produced the highest coefficient of determination ($R^2$) and least Root Mean Square of Error (RMSE) value, followed by the NIA, CP and CY (produced a negative association with ESV) (**Fig. 10**). While considering the cumulative effects of all the driving factors (except the CI factor) on multiple ESs, the highest association was observed between the food production service and driving factors, followed by waste treatment, soil formation and retention, water regulation, climate regulation, biodiversity management, gas regulation, recreation, and raw material production services (**Fig. 11**). The strong positive association between the food production and cropping factors are indicating that the GR led agrarian expansion has significantly improved the agricultural ESs of the country.

## *3.5 Limitation and future scope*

Though this study has incorporated several valuation approaches and unit values to estimate the per unit ESVs for different key ESs, still we acknowledge some limitations exist in the quantification and valuation process. The direct benefit transfer method (DBM) proposed by Costanza et al., (1997, 2014) was based on the assumption of spatial homogeneity and invariability of unit values specified for an equivalent biome. The direct linkages of existing unit values to corresponding land units without considering the local and regional landscape variability and socio-ecological diversity may produce under (or over) estimates. Apart from this, we have adopted the Xie et al., 2008 equivalent value coefficients for estimating the cropland equivalent factor which was mainly calculated for the Chinese landscape. Since we considered a country level assessment, we assumed that the equivalent weights for the selected ESs would be spatially invariant. Additionally, not only the cropped area and crop production but several other factors, i.e. the changes of cropping structure, plantation types, landscape composition and configuration etc. can also be responsible for the changes of ESVs (Cai et al., 2013; Qiu and Turner, 2015). Liu et al., (2017) study revealed that due to the changes of plantation types, an estimated 359.44 × 104 USD cropland ES has been increased, which contributed 22.97% to the total increase. Our study also indicates the most important feature of a cropland ecosystem is producing multiple key ESs, that creates natural capital. However, in our study, we have not considered the trade-offs and synergies among the major ESs due to LULC changes. This could be helpful to track overall complementary nature of many interdependent ESs. For instance, the factors (expansion of cropping area, uses of chemical fertilizer, irrigation) which are responsible for the increase of food production service in any given ecosystem was found to be detrimental for water quality and supply of fresh water services in many cases across the world (Keesstra et al., 2018; Awasthi et al., 2016). Finally, the efficacy of emerging approaches including machine learning based spatially explicit models, linear and non-linear optimization needs to be assessed under different agro-climatic and geographical conditions, before adopting it as a general solution mechanism for real-world problems. Future research will be directed in this way to resolve the methodological uncertainties and biases that exist in this valuation study.

## 4. Conclusion

This study quantifies the ESVs of different ecoregions of India from 1985 to 2005 using remote sensing based LULC products and crop production statistics. Using the five unit values, the mean total ESV (Billion US$ year-1) of India was estimated 829, 830, and 845 for 1985, 1995 and 2005, respectively. Due to the GR led agricultural expansion, the average cropland ESV has increased from 389. 32 Billion US$ year in 1985 to 402.54 Billion US$ year in 2005 (a net increase of 13.22 Billion US$ cropland ESVs during 1985–2005). The cropland has increased substantially during 1995–2005, mainly due to excess monsoon rainfall, and due to the major/minor/small irrigation programs which were launched at different times as the outcome of the GR, which significantly reformed the crop production scenarios of the country. Among the five explanatory factors, the TCA has explained the maximum model variance, followed by NIA, CP, CY, and CI. The CY was found negatively associated with ESVs, whereas, the CI is not significantly correlated with ESVs. A significant forest cover is lost during 1985-2005, mostly due to deforestation, shifting cultivation, timber and fuel-wood collection, and wildfires. The alarming rate of forest cover loss, especially in the Northeast and Himalayan states of India forms a serious environmental threat for sustainable natural resource management. Additionally, amongst the nine ESs, a strong positive association between the food production service and cropping factors indicates that the GR led agrarian expansion has significantly improved the agricultural ESs of the country. However, considering the elasticity of ESVs of the major ecosystem types of the country, wetlands, water bodies, and forest land are the most sensitive ecosystems to LULC change. Furthermore, these land uses exhibit a higher service providing capacity than any semi-artificial and artificial landscape. Therefore, land degradation prevention policies should be implemented for the reclamation of cropland from fallow land and to reduce the over-consumption of agricultural land by intensifying the cropping practices rather expanding crop area at the expense of removing forest and natural green cover. The findings of this study also provides beneficial information for farmers, agronomists, environmentalist, planners, land administrators, managers, and decision-makers for sustainable agricultural management as well as natural resources and conservation of the ecosystems of the region.

**Acknowledgement**

The authors express their sincere gratitude to the anonymous reviewers and the Editorial Board for fruitful and constructive comments to enhance the quality of paper. The authors are grateful to Ms Bhumika Uniyal for her constant support and feedback. SS and SC acknowledge University Grant Commission (UGC) for research fellowship. SB acknowledge INSPIRE Fellowship (Award Number: IF131138) by Department of Science & Technology (DST, New Delhi) for doctoral research. SR thanks the Ministry of Human Resource Development (MHRD, New Delhi) for doctoral research fellowship.

**List of Figures**

**Fig. 1** The spatial distribution of different land use/land cover categories in 2005 in India.

**Fig. 2** (a) The mean ESV (Billion US$ year) derived from the five unit values, (b) percentage contribution of ESV by seven LULC categories, and (c) changes of ESV (Billion US$) during the research period (1985 – 2005).

**Fig. 3** Coefficient of elasticity of ESV to LULC changes in India during 1985 – 1995, 1995 – 2005, and 1985 – 2005, respectively.

**Fig. 4** Temporal changes of total food grains, total crop area, and the net irrigated area in India during 1950 – 51 to 2005 – 06.

**Fig. 5** The local GWR $R^2$ approximation between the explanatory variables (total crop area, crop production, crop yield, net irrigated area, and cropping intensity) and ESV.

**Fig. 6** Normalized importance of the explanatory variables derived from ANN.

**Fig. 7** Simple linear regression model between multiple ESs.

**Fig. 8** Correlation matrix between the nine ESs (GR = Gas Regulation, CR = Climate Regulation, WR = Water Regulation, SFR = Soil Formation and Retention, WT = Waste Treatment, BDM = Biodiversity Maintenance, FP = Food Production, RMP = Raw Material Production, and RCA = Recreation, Culture, and Aesthetic), and five explanatory factors (TAC = Total Crop Area, CP = Crop Production, CY = Crop Yield, NIA = Net Irrigated Area, and CI = Cropping Intensity).

**Fig. 9** Correlation matrix between the five explanatory factors (TAC = Total Crop Area, CP = Crop Production, CY = Crop Yield, NIA = Net Irrigated Area, and CI = Cropping Intensity) and ESV.

**Fig. 10** Coefficient of Determination ($R^2$) and correlation between the driving factors and ESV.

**Fig. 11** Coefficient of Determination ($R^2$) and correlation between the 9 ESs and all driving factors.

**Fig. S1** LULC conversion during 1985 – 1995, 1995 – 2005, and 1985 – 2005, respectively in India.